\documentclass[12pt]{article}
\usepackage{amssymb}
\usepackage{epsfig}
\usepackage{graphicx}

\def\eq#1{{Eq.~(\ref{#1})}}
\def \bea {\begin{eqnarray}}
\def \eea {\end{eqnarray}}
\def \bea* {\begin{eqnarray*}}
\def \eea* {\end{eqnarray*}}

\def \be {\begin{equation}}
\def \ee {\end{equation}}
\def \bes {\begin{equation*}}
\def \ees {\end{equation*}}

\def \lm  {\lambda}

\def \pa  {\partial}

\begin{document}

\begin{titlepage}

\begin{center}

\vskip 1.5 cm {\Large \bf Conformal Symmetry, Rindler Space and The AdS/CFT Correspondence}
\vskip 1 cm {Prasant Samantray\footnote{prasant.samantray@icts.res.in}$^{a}$,
and T. Padmanabhan\footnote{nabhan@iucaa.ernet.in}$^{b}$}\\
{\vskip 0.75cm $^{a}$ International Centre for Theoretical Sciences, IISc Campus, Bengaluru 560012, India} 
{\vskip 0.75cm $^{b}$ IUCAA, Pune University Campus, Ganeshkhind, 
Pune 411007, India}

\end{center}

\vskip  .25 cm

\begin{abstract}
\baselineskip=16pt
\noindent Field theories in black hole spacetimes undergo dimensional reduction near horizon (in the Rindler limit) to two dimensional conformal field theories. We investigate this enhancement of symmetries in the context of gauge/gravity duality by considering Rindler space as boundary of Anti-de Sitter space in three spacetime dimensions. We show that the loxodromy conjugacy class of the SO(2,2) isometry group is responsible for generating the special conformal transformations on the boundary under RG flow. We use this approach to present an alternative derivation of the two-point function in Rindler space using AdS/CFT correspondence.

\end{abstract}

\end{titlepage}

\section{Introduction}
It is well known that any field theory (classical or quantum) enjoys enhanced symmetries near the horizon of a black hole. For example, interacting  field theories reduce to two dimensional free field theories near the horizon (see \cite{Carlip, TP, Solodukhin} and references therein). This dimensional reduction to two dimensional {\it conformal} field theories results in some interesting features. Work in the past couple of decades suggests that the emergence of full conformal symmetry near horizon could be the evidence for something more fundamental related to the nature of spacetime itself. A complete understanding of this feature is necessary in order to decipher the true degrees of freedom of any viable quantum theory of gravity. The observation that such a near horizon dimensional reduction might be a universal feature stems from an earlier work \cite{Strominger} that black hole entropy (in three dimensions) can be calculated exactly using the Cardy's formula \cite{Cardy}, which enumerates the density of states of a two-dimensional conformal field theory. Thus, it seems that two dimensional conformal field theory has fundamental relevance in the description of black hole microstates.

Since Rindler space appears as the near horion limit (of non-extremal black holes), study of quantum field theory in Rindler space assumes critical significance. In addition, owing to the dimensional reduction, it finally boils down to studying conformal field theory in two-dimensional Rindler space. Two dimensional Rindler space is represented by
\be
ds^2 = -\xi^2 dt^2 + d\xi^2 \nonumber 
\ee 

\noindent and has certain interesting features. The Rindler horizon is at $\xi= 0$. If we transform $\xi \rightarrow 1/\rho$, then the above metric becomes
\be
ds^2 = \frac{1}{\rho^4}\left(-\rho^2 dt^2 + d\rho^2 \right) \nonumber 
\ee 

\noindent which is just Rindler metric re-scaled by the conformal factor $1/\rho^4$. However, since we are dealing with a conformal field theory, the overall scale is irrelevant. This implies that if $\phi(t,\xi)$ is a solution to the field equation, so is $\phi(t,1/\xi)$. Since $\xi$ is just a coordinate, this feature is strongly remniscient of the UV/IR connection. In fact, similar  mathematics is involved in the modular invariance of the world sheet string theory, which makes loop amplitudes in string theory finite. 

Interestingly, the equation for a scalar field near a non-extremal black hole (more generally in the Rindler limit of the horizon) reduces to Schr$\ddot{\mathrm{o}}$dinger equation (with zero energy eigenvalue) in an inverse square potential \cite{TPSreeni}. To see this, consider an interacting scalar field with potential $V(\Phi)$ in the spacetime 

\begin{equation}
\label{metric}
ds^2 = f(r)dt^2 - f^{-1}(r)dr^2 + dL^2_{\perp}
\end{equation}

where $f(r)=0$ at $r=r_0$ with $f'(r)$ being finite and nonzero at $r_0$, and $dL^2_{\perp}$ is the transverse part of the metric. By expanding $f(r)$ near the horizon, we get  

\begin{equation}
f(r) = f'(r_0)(r - r_0) + \mathcal{O}[(r - r_0)^2] \approx f'(r_0)(r - r_0).
\end{equation}

In the case of a Schwarzschild black hole, we have $f'(r_0) = r_0^{-2}$ with $r_0 = 2M$ as the Schwarzschild radius. The field equation for the scalar field $\Phi(t,r)$ written for the metric in \eq{metric} reads

\begin{equation}
c^{-2}f(r)^{-1}\,\partial^2_t\Phi - \partial_r\left(f(r)\partial_r\Phi\right) = - V'(\Phi). 
\end{equation}

Substituting the ansatz $\Phi (r,t) = e^{-i \omega t}\frac{\psi(r)}{\sqrt{f(r)}}$ in the above equation, we see that $\psi(r)$ satisfies the equation

\begin{equation}
\label{effschr}
- \frac{\hbar^2}{2}\,\frac{d^2\psi(r)}{dr^2} - \frac{\alpha}{(r - r_0)^2}\,\psi(r) = 0    
\end{equation}

where $\alpha = \hbar^2\omega^2/2 c^2[f'(r_0)]^2$ close to the horizon. We also note that close to the horizon, the $V'(\Phi)$ term does not contribute in the leading order and (for the Schwarzschild spacetime, $\alpha = \hbar^2\omega^2r_0^2/2c^2$) $\alpha$ has dimensions of $\hbar^2$.  Setting $ x = (r - r_0) $, reduces \eq{effschr} to the Schr$\ddot{\mathrm{o}}$dinger equation for a particle in $-\tilde{\alpha}/x^2$ potential, 

\begin{equation}
\label{effschr1}
- \frac{\hbar^2}{2m}\,\frac{d^2\psi(x)}{dx^2} - \frac{\tilde{\alpha}}{x^2}\,\psi(x) = \mathcal{E}\psi(x)
\end{equation}

\noindent where $\tilde{\alpha} =\alpha/m$ and we take the limit $\mathcal{E}\,\rightarrow\,0$ at the end of the calculations. Thus the problem of scalar field in the near horizon (or Rindler limit) is mapped to the Schr$\ddot{\mathrm{o}}$dinger equation for a particle in $-\tilde{\alpha}/x^2$ potential near the origin. 

If we consider just the field equation without any reference to Schr$\ddot{\mathrm{o}}$dinger equation, i.e. when $\mathcal{E}=0$, it is clear that $x \rightarrow \lm x$ leaves \eq{effschr1} invariant. This scale invariance of the field equation is just an example of the larger conformal group that any field theory enjoys close to the horizon. Obviously, the near horizon limit (the Rindler limit) of black hole is playing a crucial role in such symmetry enhancement. In \cite{Carlip}, the classical central charge was used in Cardy's formula to calculate the entropy of black hole. The reason why Cardy's formula gives the correct gravitational entropy in three dimensions still remains a mystery. However, this result also suggests that this classical central charge would perhaps be protected from quantum corrections in any quantum theory of gravity. This is also suggestive of the possibility that trying to ``quantize" classical gravity might not be optimal. Rather, we should consider classical gravity as the low energy effective theory of some final correct quantum gravity theory. AdS/CFT correspondence \cite{Maldacena} holds promise in this regard since the CFT defines quantum gravity in AdS space. Since two dimensional Rindler space is what captures all the crucial physics in the near horizon limit of non-extremal black holes, it is but natural to try and understand these symmetries in the larger context of the AdS/CFT correspondence. This is because while quantum gravity is poorly understood in asymptotically flat spacetimes, we understand quantum gravity somewhat better in asymptomatically AdS spacetimes via the AdS/CFT correspondence.


\section{Rindler space as the boundary of AdS}
\noindent We would like to express Rindler space as the conformal boundary of AdS space in order to understand quantum field theory in Rindler space using the gauge/gravity correspondence. We would also like to map the conformal symmetry enjoyed by the field equations in Rindler space to the bulk isometries of AdS. To this effect, we restrict ourselves to three dimensional AdS space for simplicity. The boundary theory would then live on two dimensional Rindler space which is the most relevant for our purpose. This is so because as seen in the previous section, any interacting field theory near the horizon reduces to a free conformal field theory in the Rindler limit. Generalization to higher dimensions is relatively straightforward. 

We began by presenting an explicit map between the boundary conformal symmetry and the bulk isometries when the conformal boundary is just two dimensional Minkowski space. We employ the Poincare coordinatization for AdS$_3$ (with the AdS curvature scale set to unity for convenience) as
\begin{eqnarray}
X_0 &=& \frac{t}{z} \nonumber \\
X_1 &=& \frac{x}{z} \nonumber \\
X_2 &=& \frac{1}{2z} \left(z^2 - 1 + x^2 - t^2 \right) \nonumber \\
X_3 &=& \frac{1}{2z} \left(z^2 + 1 + x^2 - t^2 \right)  \label{Poincare}
\end{eqnarray} 

\noindent The AdS$_3$ metric then becomes
\begin{eqnarray}
ds^2 &=& -dX^2_0 + dX^2_1 + dX^2_2 - dX^2_3 \nonumber \\
     &=& \frac{1}{z^2} \left(dz^2 - dt^2 + dx^2 \right)  \label{AdS3}
\end{eqnarray}

\noindent There are six independent global Killing generators and SO(2,2) isometry is manifest, which of course also happens to be the symmetry group of conformal field theory in two dimensions. This matching of symmetry groups is considered as one of the justifications for the proposed gauge/gravity correspondence. We will make this justification more concrete by considering the bulk isometries and showing that full conformal symmetry of the boundary theory emerges as we consider an RG flow towards the UV (i.e. as $z \rightarrow 0$). 
\\

\noindent First, we observe that action of the operators $\pa_t$, $\pa_x$, and $t \pa_x + x\pa_t$ leave the metric in \eq{AdS3} unchanged. These generators are just isometries of two dimensional Minkowski space which also is the boundary of AdS$_3$. It is straightforward to work out the corresponding bulk AdS generators in the embedding space from \eq{Poincare} as
\begin{equation}
\pa_t \leftrightarrow \left(X_3 \pa_0 - X_0 \pa_3 \right) - \left(X_2 \pa_0 + X_0 \pa_2 \right) \nonumber
\end{equation}
\\
\be 
\pa_x \leftrightarrow \left(X_2 \pa_1 - X_1 \pa_2 \right) - \left(X_3 \pa_1 + X_1 \pa_3 \right) \nonumber
\ee
\\
\be 
t \pa_x + x\pa_t \leftrightarrow  X_0 \pa_1 + X_1 \pa_0 
\ee
\\
We also have the dilatation isometry $(z, t, x)\rightarrow (\lm z, \lm t, \lm x) $, which happens to be the Lorentz boost generator $X_2 \pa_3 + X_3 \pa_2$ in the embedding space. This is the underlying isometry relevant in the discussion of UV/IR connection in gauge/gravity duality. However, the most interesting isometry is generated by the linear combination of the two {\it{loxodromic}} generators in the embedding space. As discussed in \cite{ParikhSamErik}, AdS$_3$ has two loxodromic generators i.e.
\begin{eqnarray}
L_1 = a_1 \left(X_0 \pa_2 + X_2 \pa_0 \right) + a_2 \left(X_3 \pa_1 + X_1 \pa_3 \right) \nonumber \\
L_2 = a_1 \left(X_0 \pa_3 - X_3 \pa_0 \right) + a_2 \left(X_1 \pa_2 - X_2 \pa_1 \right)  
\end{eqnarray}
where $a_1$ and $a_2$ are real valued parameters. $L_1$ corresponds to sum of two boosts, whereas $L_2$ corresponds to sum of two rotations in the embedding space. These generators cannot be reduced to either a single boost or rotation by any symmetry transformation. Therefore, they belong to a distinct conjugacy class of SO(2,2).

It turns out that the linear combination $L_1 - L_2$ generates isometries which under the RG flow become the special conformal transformations of the boundary theory. This can be seen by the action of $L_1 - L_2$ on the embedding coordinates. Under infinitesimal transformations, the new coordinates become
\begin{eqnarray}
\bar{X_0} &=& X_0 + a_1 \left(X_2 + X_3\right) \nonumber \\
\bar{X_1} &=& X_1 + a_2 \left(X_2 + X_3\right) \nonumber \\
\bar{X_2} &=& X_2 + a_1 X_0 - a_2 X_1 \nonumber \\
\bar{X_3} &=& X_3 - a_1 X_0 + a_2 X_1  
\end{eqnarray}

Or, in terms of the Poincare coordinates in \eq{Poincare}
\begin{eqnarray}
\bar{z} &=& \frac{z}{1 - 2a_1 t + 2 a_2 x} \nonumber \\
\bar{t} &=& \frac{t + a_1 \left(z^2 + x^2 - t^2 \right)}{1 - 2a_1 t + 2 a_2 x} \nonumber \\
\bar{z} &=& \frac{x + a_2 \left(z^2 + x^2 - t^2 \right)}{1 - 2a_1 t + 2 a_2 x} \end{eqnarray}

\noindent We see that when $z \rightarrow 0$, the boundary Poincare coordinates transform as
\be
\bar{x}^\mu = \frac{x^\mu + a^\mu (x\cdot x)}{1 + 2 (a\cdot x)} \approx x^\mu - 2 (a\cdot x)x^\mu + a^\mu (x\cdot x) 
\ee
where $x^\mu = (t,x)$ and $a^\mu = (a_1,a_2)$. These, of course are the infinitesimal form of special conformal transformations. Therefore, we see that the loxodromic conjugacy class of SO(2,2) generates special conformal algebra at the boundary (as to why only the specific linear combination, $L_1 - L_2$, works, is not very clear at present and warrants further investigation). 

After having understood the emergence of boundary conformal symmetries from the bulk isometries of AdS$_3$, we can now address the origin of the scaling symmetry associated with the field equation near the horizon of an non-extremal black hole.  Since the near horizon geometry in this case is Rindler space, we express AdS$_3$ in coordinates where Rindler space appears as its boundary. The metric reads

\be
\frac{1}{z^2} \left(dz^2 - \xi^2 d\tau^2 + d\xi^2 \right) \label{RAdS}
\ee
where the new coordinates are related to our earlier Poincare coordinates in \eq{Poincare} as 
\begin{eqnarray}
t = \xi \sinh \tau \nonumber \\ 
x = \xi \cosh \tau \label{RCoordinates}
\end{eqnarray}

\noindent With these transformations, the boundary theory now has just four independent symmetry generators instead of six (since $\pa_t$ and $\pa_x$ are no longer isometries). The scaling symmetry of the field equation then is the consequence of the transformation $\xi \rightarrow \lm \xi$, which we now understand is just the result of the RG flow of the Lorentz boost generator $X_2 \pa_3 + X_3 \pa_2$ towards the conformal boundary. 

\section{Rindler correlation functions from AdS/CFT}
We now turn to the boundary field theory which lives on two dimensional Rindler space. The central object in any theory is the two-point function characterizing various correlations. In this section we present an alternative and simpler derivation of the Rindler two-point functions using the AdS/CFT correspondence.  We start with the Wick rotated version of the metric in \eq{RAdS}

\be
ds^2 = \frac{1}{z^2} \left(dz^2 + \xi^2 d\phi^2 + d\xi^2 \right) \label{RAdSW}
\ee
where $i\phi = \tau$, and proceed according to \cite{GKP,WittenHolography}.

Our basic tool is the formula relating the bulk and the boundary dynamics
\be
\int_{\rm bulk} {\cal D} \phi \, e^{iI[\phi]} = \left \langle e^{\int \!  \phi_0 {\cal O}} \right \rangle_{CFT} \; , \label{AdS-CFT}
\ee 
where $\phi_0$ is the boundary value of the field $\phi$, and ${\cal O}$ is the conformal operator which couples to $\phi_0$ at the boundary. Using the supergravity approximation (large $N$ and large 't Hooft coupling), we can approximate the bulk partition function by its saddle-point
\be
\int_{\rm bulk} {\cal D} \phi \, e^{iI[\phi]} \sim e^{i I [\phi_{\rm cl}]} \; , \label{Sugra}
\ee 
where $I [\phi_{\rm cl}]$ is the classical bulk action. In order to evaluate the bulk action, we need to first solve for $\phi_{\rm cl}$. Considering a massless bulk scalar field for simplicity, the action is
\be
I[\phi] = \int \left ( -\frac{1}{2} \left(\partial \phi \right)^2 \right ) \frac{\xi}{z^3} dz d\xi d\phi \; .\label{action}
\ee 

\noindent Integrating \eq{action} by parts, and separating the bulk and the surface terms, we get for the variation of the action
\be
\delta I[\phi_{\rm cl}] \sim \int g^{\mu \nu} \delta \phi_{\rm cl} \, \partial_{\mu}\phi_{\rm cl} \, d\Sigma_\nu \; , \label{boundaryterm}
\ee
where $d\Sigma_\nu$ is the surface normal to the bulk $z$ coordinate. The variation of the bulk term vanishes on-shell owing to the equations of motion. Since we wish to evauate this action at the boundary, we impose an infrared cut-off close to the boundary at $z = \epsilon$ in order to regulate the action.

We can write the solution for the equation of motion of the bulk field as
\be
\phi_{\rm cl} (z, \xi, \phi) = \sum^{n=\infty}_{n=-\infty} \int^{\infty}_{0} \lm_{k,n} \frac{z K_1 (k z)}{\epsilon K_1 (k \epsilon)} J_n (k\xi) e^{in\phi} \sqrt{k} dk \label{bulk-field} 
\ee
where $K_1 (kz)$ is the modified Bessel function, and $\lm_{k,n}$ is the coefficient of the mode expansion. The choice of this particular solution is based on the fact that $K_1 (kz)$ and $J_n (k\xi)$ are both regular as $z \rightarrow \infty$ (Poincare horizon), and $\xi \rightarrow 0$ (Rindler horizon), respectively. Therefore, we choose regularity at the horizons as the criteria for our choice of the solution. The boundary two-point function is derived by twice differentiating the boundary term of action in \eq{boundaryterm}. Inserting \eq{bulk-field} into \eq{boundaryterm}, and differentiang twice with respect to $\lm$, we get (neglecting the unimportant prefactors)

\be
\langle {\cal O}(k,n) {\cal O}(q,m)\rangle \sim (-1)^n \delta_{m+n,0}~k^2 \log {k}  \label{Fluxfactor}
\ee

\noindent Using the Jacobi-Anger expansion \cite{Stegun} and Fourier transforming the above two-point function to position space, we get
\begin{eqnarray}
\langle {\cal O}(\xi_1,\phi_1) {\cal O}(\xi_2,\phi_2)\rangle &=& \sum^{n=\infty}_{n=-\infty} \int k^2 \log {k}~ J_n (k\xi_1) J_n (k\xi_2) e^{in(\phi_1 - \phi_2)} k dk \nonumber \\
&=& \frac{1}{\left[\xi^2_1 + \xi^2_2 - 2 \xi_1 \xi_2 \cos (\phi_2 - \phi_1)\right]^2} \label{Two-pointfnWick}
\end{eqnarray}

\noindent Wick rotating back to real time, we get
\be
\langle {\cal O}(\xi_1,t_1) {\cal O}(\xi_2,t_2)\rangle \sim \frac{1}{\left[\xi^2_1 + \xi^2_2 - 2 \xi_1 \xi_2 \cosh (\tau_2 - \tau_1)\right]^2} \label{Two-pointfn} 
\ee

\noindent Since the boundary theory is a CFT, as expected, the the two-point function is typically of the form $|x - y|^{-\Delta}$, where $\Delta$ is the conformal weight of the operator. Comparision with the above expression tells us that the conformal weight of ${\cal O}$ is $\Delta = 2$. An example of such an operator is $\pa \phi$ (i.e. if $\phi$ is a scalar field living in two dimensional Rindler space). This is the result for a massless bulk scalar field. In the case of a massive bulk scalar field in AdS$_3$, the solution for the bulk field can be written as
\be
\phi_{\rm cl} (z, \xi, \phi) = \sum^{n=\infty}_{n=-\infty} \int^{\infty}_{0} \lm_{k,n} \frac{z K_{\nu} (k z)}{\epsilon K_{\nu} (k \epsilon)} J_n (k\xi) e^{in\phi} \sqrt{k} dk \label{bulk-field-massive} 
\ee

\noindent where $\nu = (1 + \sqrt{1 + m^2})/2$. Following the previous analysis for the computation of two-point function, we get

\begin{eqnarray}
\langle {\cal O}(\xi_1,\phi_1) {\cal O}(\xi_2,\phi_2)\rangle 
= \frac{1}{\left[\xi^2_1 + \xi^2_2 - 2 \xi_1 \xi_2 \cos (\phi_2 - \phi_1)\right]^\Delta} \label{Two-pointfnWick-massive}
\end{eqnarray}

\noindent where $\Delta = 1 + \sqrt{1 + m^2}$ is the conformal weight in this case, and $m$ is the mass of the field. This of course is completely consistent with the standard dictionary of AdS/CFT where the conformal weight is related to the mass of the bulk field being considered (in the supergravity approximation). The two-point functions in \eq{Two-pointfnWick} and \eq{Two-pointfnWick-massive} reflect the correct periodicity ($=2\pi$) in imaginary time, thereby producing the correct Rindler temperature. Generalization to higher dimensional Rindler spaces is straightforward.

In the foliation that we considered, Rindler space appeared as the boundary of AdS. Calculation of the Hawking-Bekenstein entropy for the Rindler horizon in \eq{RAdS} gives (Rindler horizon is at $\xi=0$)
\be
S_{BH} = \frac{1}{4G_3} \int \frac{dz}{z} \label{Entropy-BH} 
\ee

\noindent which of course is divergent. Since by the AdS/CFT correspondence, the partition functions of the bulk and boundary theories are equal, we can use the Cardy formula to calculate the entropy of the boundary CFT. The boundary is at the surface $z = z_0 \rightarrow 0$. The boundary metric then reads
\be
ds^2_{b} = \frac{\xi^2}{z^2_{0}} \left(-d\tau^2 + \frac{d\xi^2}{\xi^2} \right) 
\ee 
Since the boundary theory is a CFT, it is insensitive to the overall conformal factor $\frac{\xi^2}{z^2_{0}}$. Therefore, the entropy of the CFT is given by \cite{Cardy}
\be
S_{CFT} = \frac{\pi}{3 \beta} c ~ {\rm V}= \frac{\pi}{3} \frac{3}{2 G_3} \frac{1}{2 \pi} \int \frac{d\xi}{\xi} = \frac{1}{4G_3} \int \frac{d\xi}{\xi}  \label{Entropy-CFT} 
\ee
where ${\rm V}$ is the spatial volume of the boundary, and $c = 3/2 G_3$ is the central charge of the unitary CFT as calculated by Brown and Henneaux \cite{BrownHenneaux} in units where the AdS scale has been set to unity. As expected, the CFT entropy in \eq{Entropy-CFT} is divergent as well. However, it is remarkable that both $S_{BH}$ and $S_{CFT}$ have the correct entropy density ($=1/4G_3$). Also, since both $S_{BH}$ and $S_{CFT}$ are divergent, we can always chose a regularization scheme such that $\int dz/z \sim \int d\xi/ \xi$. Therefore, the boundary CFT correctly predicts the entropy of the Rindler horizon (Note however that in \cite{ParikhSamantray}, it was found that a simple free field theoretic calculation yields an $S_{CFT}$ which scales differently than $S_{BH}$ in a specific foliation of AdS where the boundary CFT lives on de Sitter space).

Alternatively, one can foliate AdS$_3$ in a way such that the Rindler horizon is in the bulk, and the boundary metric is just two dimensional flat Minkowski space \cite{ParikhSamantray}. It would be interesting to map the states of a quantum field on the Rindler horizon (or on the stretched horizon) to the operators of the boundary conformal field theory. Other physical quantities of interest, like the correlations of the boundary stress tensor in Rindler space can also be calculated by considering perturbations of the bulk metric, and applying the formula \eq{AdS-CFT}. The advantage of using gauge/gravity duality to compute such quantities in Rindler space is that the correlation functions of operators like $T_{\mu\nu}, \pa \phi$ etc. are automatically regularized (this is related to the infrared cut-off imposed in regulating the bulk path integral in \eq{AdS-CFT}).

\section*{Acknowledgements} 
P.S thanks IUCAA for the hospitality where this work was done. TP is partially supported by the J.C.Bose grant of Department of Science and Technology, India.


\end{document}